\documentclass[12pt]{article}
\pdfoutput=1
\usepackage{amsmath,amssymb,amsfonts,epsfig,graphicx,euscript,float}%
\usepackage[pdftex,bookmarks,bookmarksnumbered,linktocpage,pdfstartview=FitH]{hyperref}
\hypersetup{colorlinks,%
citecolor=red,%
filecolor=blue,%
linkcolor=blue,%
urlcolor=blue,%
pdftex}
\usepackage[all]{hypcap}
\numberwithin{equation}{section}
\usepackage{cite}
\newcommand{\be}{\begin{equation}}
\newcommand{\ee}{\end{equation}}
\newcommand{\bea}{\begin{eqnarray}}
\newcommand{\eea}{\end{eqnarray}}
\providecommand{\keywords}[1]{\textbf{Keywords}: #1}
\begin{document}
\hfill%
\vspace{0.5cm}
\begin{center}
{ \Large{\textbf{Meson Thermalization by Baryon Injection in D4/D6 Model}}} 
\vspace*{1.2cm}
\begin{center}
{\bf Z. Rezaei$^{a,b,}$\footnote{z.rezaei@aut.ac.ir}}%

\vspace*{0.3cm}
{\it {${}^a$}Department of Physics, Tafresh University, Tafresh 39518 79611, Iran}\\
{\it {${}^b$}School of Particles and Accelerators, Institute for Research in Fundamental Sciences (IPM), P.O. Box 19395-5531, Tehran, Iran}\\

\vspace*{0.8cm}
\end{center}
\end{center}

\bigskip
\begin{center}
\textbf{Abstract}
\end{center}
We study meson thermalization in a strongly coupled plasma of quarks and gluons using AdS/CFT duality technique. Four dimensional large-N$_c$ QCD is considered as a theory governing this quark-gluon plasma (QGP) and D4/D6-brane model is chosen to be its holographic dual theory. In order to investigate meson thermalization, we consider a time-dependent change of baryon number chemical potential. Thermalization in gauge theory side corresponds to horizon formation on the probe flavor brane in the gravity side. The gravitational dual theory is compactified on a circle that the inverse of its radius is proportional to energy scale of dual gauge theory. It is seen that increase of this energy scale results in thermalization time dilation. 
In addition we study the effect of magnetic field on meson thermalization. It will be seen that magnetic field also prolongs thermalization process by making mesons more stable.

\keywords{Gauge/gravity duality, Thermalization}

\newpage

\tableofcontents

\section{Introduction}

Quak-gluon plasma (QGP) produced in heavy ion collisions in RHIC and
LHC is believed to be a strongly coupled plasma
\cite{Shuryak:2003xe,Shuryak:2004cy}. Since QGP is totally out of
equilibrium immediately after collision, it is difficult to study
it. After a very short time ($\tau \leq 1 fm/c$) the hydrodynamic
regime  governs the plasma which shows a local equilibrium
\cite{Heinz:2004pj,Luzum:2008cw}. This very fast process is called
rapid thermalization. Perturbative methods are not suitable to study
such a system due to being strongly coupled. Among non-perturbative
methods AdS/CFT correspondence has been recently successful to
describe rapid thermalization.

AdS/CFT correspondence or gauge/gravity duality is generally a
duality between a strongly coupled field theory and a classical
gravitational theory
\cite{Maldacena:1997re,Gubser:1998bc,Witten:1998qj,CasalderreySolana:2011us}.
In its original form, AdS$_5$/CFT$_4$ duality introduces a
correspondence between a four dimensional super conformal Yang-Mills
theory with the gauge group of SU(N$_c$) and type IIB superstring
theory in AdS$_5\times$S$^5$ geometry. This duality is more useful
in the limit of large-N$_c$ and large t'Hooft coupling $\lambda$. In
this limit a classical type IIb supergravity theory in
AdS$_5\times$S$^5$ geometry will be dual to a strongly coupled super
Yang-Mills (SYM) theory that one can not deal with it
perturbatively. There are several generalizations to this original
correspondence. For instance if one is interested in thermal SYM
theory the dual theory is a supergravity in AdS-Schwarzschild black
hole background \cite{Witten:1998zw}. In this case the temperature
of SYM theory is Hawking temperature of the black hole. Also we can
add matter fields to the set up as another generalization
\cite{Karch:2002sh}. In the original form of AdS/CFT duality only
gauge fields are present in SYM theory and in order to add matter
fields (quarks) in the fundamental representation one should add
probe flavor D-branes to the gravity side. In  AdS$_5$/CFT$_4$  case
that D3-branes form AdS$_5\times$S$^5$ geometry, the probe limit
means that the number of flavor D7-branes should be much less than
the number of D3-branes. In this set up open strings stretching
between D3 and D7-branes represent quarks in SYM theory while open
strings with both ends on D7-branes are considered as meson degrees
of freedom.

There is another model for the gravity dual theory of a four
dimensional pure SU(N$_c$) Yang-Mills theory proposed by Witten
\cite{Witten:1998zw}. In this model the background geometry is made
of D4-branes compactified on a circle of radius $M_{\rm kk}^{-1}$
\cite{Kruczenski:2003uq}. D4-brane theory before compactification is a
supersymmetric SU(N$_c$) gauge theory whose fields are fermions, scalars and gauge fields in the adjoint representation of SU(N$_c$).
The theory is effectively four dimensional at energies much less
than compactification scale $M_{\rm kk}$. Also non-periodic conditions
are applied to fermions to break all supersymmetries. Therefore this
theory is very similar to low-energy QCD of course if we add
fundamental quarks in addition to gluons by introducing probe
D6-branes.

In the gauge theory side of this set up there is a supersymmetric
five dimensional SU(N$_c$) theory coupled to a four dimensional
defect. Now if we compactify the 4-direction with the period
$2\pi/M_{\rm kk}$ and apply non-periodic conditions to D4-brane theory fermions, all
supersymmetries are broken and an effective four dimensional theory
remains at low energies $E \ll M_{\rm kk}$. In this limit scalars and
fermions in the adjoint representation become massive and decouple.
So in low energy we are left with our desirable theory that is a
four dimensional SU(N$_c$) gauge theory with a energy scale and
coupled to fundamental quarks. In this paper we are going to study rapid thermalization in such a large-N$_c$ QCD that its holographic dual is a
D4/D6 system. Note that in \cite{Ishii:2015gia} the effect of confinment on thermalization of holographic gauge theories is studied. But the dual gravitational theory differs from what we consider here and belongs to a class of Einstein-dilaton theories. They have also introduced a different procedure for throwing the system out of equilibrium. They perturb the system by specifying time dependent boundary conditions on the scalar and solve the fully backreacted Einstein-dilaton equation of motion.

Thermalization process is generally defined as
a non-adiabatic increase in temperature which drives the system out
of equilibrium and becomes possible by energy injection. In gauge/gravity
duality framework energy injection to gauge theory is done by
turning on a time dependent source which is dual to horizon
formation in bulk \cite{Chesler:2008hg, Chesler:2009cy,
Bhattacharyya:2009uu}. This class of thermalization is called gluon
sector thermalization which corresponds to black hole formation in the
bulk. Meson sector thermalization is another class of thermalization
that is manifested in horizon formation on the probe D-brane
\cite{Hashimoto:2010wv, Ali-Akbari:2013tca, Das:2010yw,AliAkbari:2012hb}. We are
interested in meson sector thermalization in this paper and we
follow the idea of \cite{Hashimoto:2010wv} for the required energy
injection. The main idea is considering a sudden change in
baryon-number chemical potential to provide the condition of heavy
ion collisions. The gravity dual to this change of baryon chemical
potential is throwing open strings into bulk from the boundary.
Since the end point of open string is a source for gauge field, the
process of throwing open strings provides a time dependent gauge
field configuration that leads to a time dependent metric for
mesons. Therefore apparent horizon formation becomes possible on the
flavor D-brane which corresponds to meson thermalization in the dual
strongly coupled field theory. Thermalization of meson degrees of
freedom means their transformation to quark and gluon degrees of
freedom in thermal equilibrium \cite{Hashimoto:2010wv}. Time scale
of thermalization is identified with the time of apparent horizon
formation from the boundary observer point of view.

Produced QGP in heavy ion collision is affected by a large magnetic field \cite{Kharzeev:2007jp} and so there are several studies of different aspects and examples of this effect \cite{Bali:2011qj,Fukushima:2008xe,Kharzeev:2010gd,Mamo:2015dea}. It seems that megnetic field can be potentially important in thermalization in QGP because thermalization occurs at early stages of heavy ion collision where the magnetic field is strong. This investigation has been the subject of some studies based on AdS/CFT duality \cite{AliAkbari:2012vt,Mamo:2015aia,Fuini:2015hba}. Here we aim to study thermalization in a large-N$_c$ QCD by utilizing a D4/D6 gravity model and observe how the scale of the theory and magnetic field affect this process. Note that present work differs from aforementioned studies by the considered gravity dual theory, D4/D6 model, and/or the method of energy injection, sudden change of baryon number chemical potential. We will show that the scale in our theory and the magnetic field both prolong meson thermalization process. This result is justified by meson binding energy considerations. 

This paper is organized as follows. In section 2 we first give a brief review of D4/D6 model in AdS/CFT duality framework. Then we show how the scale of the theory, $u_{\rm kk}$, modifies shape of the flavor D6-brane. We continue by turning on an external magnetic field and observe its effect on D6-brane configuration, too. The procedure of meson sector thermalization in D4/D6 system is addressed in section 3. We explain the mechanism of baryon injection and show how it can lead to thermalization which is dual to apparent horizon formation on the probe D6-brane. Sections 4 and 5 are devoted to numerical results and summary of the work, respectively.

\section{D4/D6 Model in the Presence of External Magnetic Field}

In this section we first give a brief review of D4/D6 model as
a confining model for AdS/CFT duality \cite{Kruczenski:2003uq}. Then
we turn on the magnetic field and study its effect on the shape of
the flavor D6-brane.

\subsection{Review of the D4/D6 Model}

The near horizon limit of a configuration of $N_c$ D4-branes
wrapping a circle in the $\tau$ direction with anti-periodic
boundary conditions for fermions is dual to type IIa supergravity
and is given by the following background
\bea\label{background} %
ds^2=\left(\frac{u}{R}\right)^{3/2}(\eta_{\mu\nu}dx^\mu
dx^\nu+f(u)d\tau^2)+\left(\frac{R}{u}\right)^{3/2}\frac{du^2}{f(u)}+R^{3/2}u^{1/2}{ds}^2_{S^4},
\eea %
\bea\label{definition} %
e^\phi=g_s\left(\frac{u}{R}\right)^{3/4},~~~~~~~~f(u)=1-\frac{u_{\rm kk}^3}{u^3}.
\eea %
The coordinates $x^\mu={x^0,...,x^3}$ are four non-compact
directions along the D4-brane while $\tau$ shows the direction of
the branes' compactification. ${ds}^2_{S^4}$ is the SO(5)-invariant
line element and $u$ is a radial coordinate in the 56789-directions
transverse to the D4-branes. The radial coordinate $u$ is known to
be bounded from below by the condition $u\geq u_{\rm kk}$. The
periodicity of $\tau$ is given by
\bea\label{compact} %
\delta\tau=\frac{4\pi}{3}\frac{R^{3/2}}{u_{\rm kk}^{1/2}},
\eea %
to avoid a conical singularity at $u=u_{\rm kk}$. In a general Dp-brane background, the parameter $R$ in \eqref{background} is
defined in terms of $N_c$, number of Dp-branes, $g_s$,
string coupling constant, and $l_s$, string length,
\be\label{general-R} %
R^{7-p}=(4\pi)^{\frac{5-p}{2}}\Gamma(\frac{7-p}{2})g_s N_c l_s^{7-p},
\ee %
where for $p=4$ reduces to
\bea\label{radius} %
R^3=\pi g_s N_c l_s^3.
\eea %

The dual $SU(N_c)$ field theory is defined by the compactification
scale, $M_{\rm kk}$, below which the theory is effectively
four-dimensional with the coupling constant $g_{\rm YM}$. These are
related to the string theory parameters by
\bea\label{scale} %
M_{\rm kk}=\frac{3}{2}\frac{u_{\rm kk}^{1/2}}{R^{3/2}}=
\frac{3}{2\sqrt{\pi}}\frac{u_{\rm kk}^{1/2}}{(g_sN_c)^{1/2}l_s^{3/2}},~~~~~
g_{\rm YM}^2=3\sqrt{\pi}\left(\frac{g_s u_{\rm kk}}{N_c l_s}\right)^{1/2}.
\eea %

The string tension $T$, the Regge slope $\alpha'$ and $l_s$ are
related by $T=\frac{1}{2\pi\alpha'}=\frac{1}{2\pi l_s^2}$. We are
interested in a background in which one can study the dual field
theory using only classical supergravity. For this reason we require
that the curvature is everywhere small compared to the string
tension. So the maximum curvature that occurs at $u=u_{\rm kk}$ and is
proportional to $(u_{\rm kk} R^3)^{-1/2}$ has
to satisfy $(u_{\rm kk} R^3)^{-1/2} \ll l_s^{-2}$. Therefore using
$\lambda=g_{\rm YM}^2 N_c$ as well as \eqref{radius} and \eqref{scale}
we obtain the following condition
\be\label{small-curv} %
\frac{u_{\rm kk}^{1/2} R^{3/2}}{l_s
^2}\approx \lambda \gg1.
\ee %
So demanding small curvatures gives rise to
large t'Hooft coupling condition in the effective four dimensional
dual gauge theory as is essential in AdS/CFT correspondence
\cite{Maldacena:1997re}. We are also interested in small local
string coupling, $g_s$, \cite{Kruczenski:2003uq} that along with
\eqref{small-curv} leads to the following condition
\bea\label{condition} %
g_{\rm YM}^4\ll\frac{1}{g_{\rm YM}^2N_c}\ll1.
\eea %

One of the most famous holographic QCD models is the D4/D6 model
\cite{Kruczenski:2003uq}. The model contains $N_c$ number of D4-branes and $N_f$ flavor D6-branes
whose configuration is given according to Table ~\ref{embedding}.
\begin{table}[H]%
\begin{center}%
\begin{tabular}{ | l | l | l | l | l | l | l | l | l | l | l |}
   \hline
    &  \multicolumn{4}{ |c| }{Boundary} & $S^1$ & radius of $S^2$ & \multicolumn{2}{ |c| }{$S^2$} & \multicolumn{2}{ |c| }{$D6_\perp$} \\ \hline
  & 0 & 1 & 2 & 3 & \multicolumn{1}{|c|}{$\tau$} & \multicolumn{1}{|c|}{$\rho$} & $\psi_1$ & $\psi_2$ & $\sigma$ & $\Phi$
  \\ \hline
 D4 & $\bullet$ & $\bullet$ & $\bullet$ & $\bullet$ & \multicolumn{1}{|c|}{$\bullet$} &  &  &  &  &
  \\
  \hline
 D6 & $\bullet$ & $\bullet$ & $\bullet$ & $\bullet$ &  & \multicolumn{1}{|c|}{$\bullet$} & \multicolumn{1}{|c|}{$\bullet$} & \multicolumn{1}{|c|}{$\bullet$}  &  & \\ \hline
   \end{tabular}
\end{center}
\caption{D6-brane configuraion in the background of D4-branes}\label{embedding}
\end{table}%
$S^4$ sphere in \eqref{background} is divided into two parts, spherical coordinates $\rho$
and $\Omega_2\equiv(\psi_1,\psi_2)$ and polar coordinates $\sigma$
and $\Phi$ which are perpendicular to D6-brane. By defining a new
radial coordinate $r^2=\rho^2+\sigma^2$ related to $u$ by
$u(r)=r\left(1+\frac{u_{\rm kk}^3}{4r^3} \right)^{2/3}$ and a new
function $K(r)=\frac{R^{3/2}u^{1/2}}{r^2}$, D4-brane metric \eqref{background} is
simply written in the form 
\bea % 
ds^2=\left(
\frac{u}{R}\right)^{3/2}\left(\eta_{\mu\nu}dx^{\mu}dx^{\nu}+f(u)d\tau^2\right)+K(r)\left(d\rho^2+\rho^2
d\Omega_2^2+d\sigma^2+\sigma^2 d\Phi^2 \right). 
\eea %
As it is
obvious from Table~\ref{embedding}, worldvolume coordinates of D6-brane are
$x^{\mu}, \rho$ and $\Omega_2$. D6-brane position in 89-plane is
determined by $\sigma=\sigma(\rho)$, $\Phi=\Phi_0$ and $\tau=\rm
constant$. Therefore the induced metric on D6-brane, $g_{ab}$, with the above ansatz can be written
as 
\bea %
ds^2(g)=\left(\frac{u}{R}\right)^{3/2}\eta_{\mu\nu}dx^{\mu}dx^{\nu}+K(r)\left[(1+\dot{\sigma}^2)d\rho^2+\rho^2
d\Omega_2^2\right] 
\eea %
where $\dot{\sigma}=\partial_{\rho}\sigma$.

\subsection{External Magnetic Field}
\begin{figure}[ht]
\begin{center}
\includegraphics[width=2.6 in]{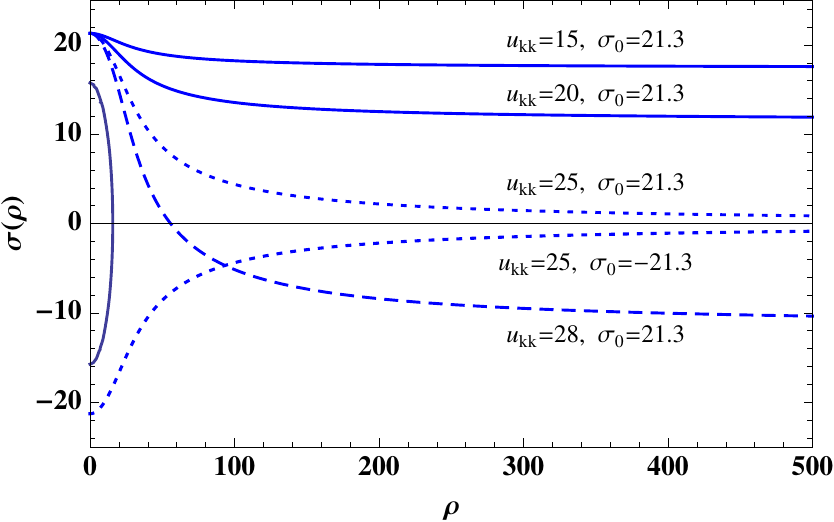}
\includegraphics[width=2.6 in]{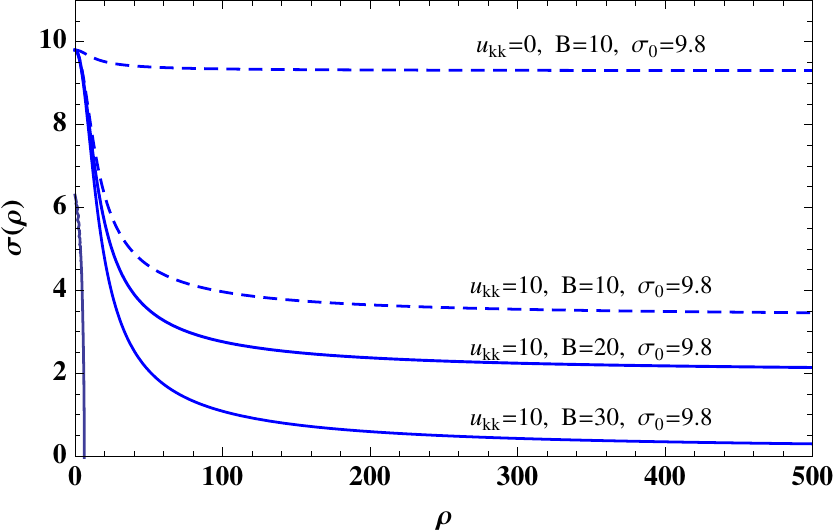}
\caption{\small D6-brane embedding for several values of $u_{\rm kk}$
(left) and $B$ (right). The circle in the center is the region $u <
u_{\rm kk}$ and is not part of the space.} \label{sigma-rho}
\end{center}
\end{figure}

In this subsection we are interested in studying the shape of
D6-brane including the effect of $u_{\rm kk}$ and external
magnetic field, $B$. To do this one should start from the DBI action of D6-brane
\be\label{DBI} %
S_{\rm DBI}=-{\cal{T}}_6 \int d^7 \xi
\sqrt{-\det (g_{ab}+2 \pi \alpha' F_{ab})},
\ee %
where ${\cal{T}}_6=\frac{2 \pi}{g_s (2\pi
l_s)^7}$ is the D6-brane tension and $\xi$'s are its worldvolume coordinates (i.e. $\xi \equiv (x^0,x^1, x^2, x^3, \rho, \psi_1, \psi_2)$ according to Table~\ref{embedding}). We consider the magnetic field as $B\equiv F_{x^1 x^2}$ and rewrite the action \eqref{DBI} as
\bea \label{D6-action} %
S_{\rm D6}=-{\cal{T}}_6 \int d^7 \xi \; \rho^2 \; \sqrt{1+\dot{\sigma}^2}\sqrt{1+(2\pi\alpha'B)^2(R/u)^{3}}\left(1+\frac{u_{\rm kk}^3}{4r^3}\right)^2.
\eea %
We need to solve the equation of motion for $\sigma(\rho)$ to study the shape of the
D6-brane. When $\rho \rightarrow \infty$ this
equation of motion finds the simple form of
\bea %
\partial_{\rho}(\rho^2\dot{\sigma})\approx -\frac{3}{2}u_{\rm kk}^3\frac{\sigma}{\rho^3}-\frac{3(2\pi \alpha'B)^2 R^3}{\rho^3}\sigma,
\eea %
which is in agreement with \cite{Kruczenski:2003uq} when $B$ is set
to zero. We solve the full equation of motion for $\sigma(\rho)$
numerically, with the regularity condition of $\sigma(\rho)$ at
$\rho=0$, (i.e. $\dot{\sigma}(\rho)|_{\rho=0}=0$) to get Fig.
\ref{sigma-rho} for zero (left) and non-zero (right) magnetic field.
If $u_{\rm kk}=0$ then $\sigma(\rho)=0$ is a solution which is
physically sensible since it is a supersymmetric case. But note that
this is only a formal argument because in the limit
$u_{\rm kk}\rightarrow 0$ the curvature at $u=u_{\rm kk}$ diverges and the
supergravity description is unreliable \cite{Itzhaki:1998dd}. When
$u_{\rm kk}\neq 0$ the force on the D6-brane does not vanish anymore and
the repulsion by the D4-branes causes the D6-brane to bend outward, Fig. \ref{sigma-rho} (left). The semicircle in this
figure is the region that $u<u_{\rm kk}$ and is not part of the space.

For large $\rho$ the $\sigma(\rho)$ expansion is given by
\be\label{sigma-exp} %
\sigma(\rho)\simeq \sigma_{\infty}
+\frac{c}{\rho},
\ee %
where according to AdS/CFT dictionary,
$\sigma_\infty$ (the value of $\sigma(\rho)$ at infinity) and $c$ correspond to bare quark mass ($m_q$) and
chiral condensate ($\langle \bar{\psi}\psi\rangle$) respectively
\cite{Karch:2006bv,Kruczenski:2003uq}
\be\label{mass-chiral}%
m_q=\frac{u_{\rm kk} \;
\sigma_{\infty}}{2\pi l_s^2},\;\;\;\;\; \langle \bar{\psi}\psi
\rangle \simeq g_{\rm YM}^2 N_C^2 M_{\rm kk}^3 c.
\ee %
For each value of
$\sigma_\infty$ there are two regular solutions which have $c$'s of
opposite sign. Such
solutions are shown in Fig. \ref{sigma-rho} (left) as dotted curves. They have the same value of $\sigma_\infty$ (equivalent to quark
mass) but the one with positive $\sigma(0)$ has lower energy and is
preferable. The dashed curve implies that by increasing $u_{\rm kk}$ there is a sever bending of D6-brane that
results in negative mass and makes the system unphysical. When we put $u(r)=u_{\rm kk}$, a circle equation $\sqrt{\rho^2+\sigma (\rho)^2}=2^{-2/3} u_{\rm kk}$ is obtained that is plotted in Fig. \ref{sigma-rho} and the interior of this circle is not part of the space.

In the right plot of Fig. \ref{sigma-rho} the effect of the magnetic
field on the shape of the probe D6-brane is demonstrated. As we
expect \cite{Ali-Akbari:2014gia}, the magnetic field also causes the
brane to bend. It is seen that for a fixed value of $B$ as one
increases $u_{\rm kk}$ the quark mass ($\sigma_{\infty}$) decreases, dashed curves in Fig.
\ref{sigma-rho} (right). In addition by increasing $B$ for a constant
value of $u_{\rm kk}$ the quark mass will decrease too, solid curves in Fig.
\ref{sigma-rho} (right). For each value of $u_{\rm kk}$ the magnetic
field cannot exceed a specific value because the mass will become
negative then and the system is not physical.

\section{Meson Thermalization in D4/D6 System}
As was previously mentioned adding probe D6-branes into the background made by D4-branes
introduces quarks to the dual gauge theory. The fluctuations of various (scalar or vector) fields living
on the D6-brane correspond to different kinds (scalar or vector) of
mesons. In this section we derive the action governing these fluctuations from DBI action of D6-brane. To study meson thermalization we will investigate the formation of apparent horizon on the D6-brane where fluctuations (mesons) live. In order to have thermalization one needs to inject energy into the system. For energy injection we follow the procedure of \cite{Hashimoto:2010wv} in which the necessary energy is provided by baryon injection.

The distance between D6 and D4-branes at $\rho \rightarrow \infty$,
($\sigma_{\infty}$), shows the quark mass according to
\eqref{sigma-exp}. Note that in this paper we put this mass equal to zero for simplicity. 

\subsection{Baryon Injection}
In the context of gauge/gravity duality baryon injection into the
strongly coupled gauge theory is done by throwing baryonically
charged open strings from the boundary onto the flavor brane. In this
framework the baryon chemical potential, $\mu$, is given by the time
component of the gauge field
\be\label{chem-pot} %
\mu=\int_0^\infty
d\rho \;\;\partial_\rho A_t(\rho).
\ee %
This time dependent chemical
potential becomes possible by introducing a source term to DBI action as
\be\label{source-general} %
S_{\rm current}=\int d^{q+1} \xi \sqrt{-g} A_a
J^a,
\ee %
where $J^a$'s are time dependent currents. In the case of D6-brane
this action will be
\be\label{source}%
\delta S={\cal{T}}_6 \;V_3\;
V(\Omega_2) \int dt\; d\rho\; (A_t j^t + A_{\rho} j^{\rho} ),
\ee %
where
$\rho^2 u^{15/4} r^{-3} R^{-3/4} \sqrt{1+\dot{\sigma}^2} ~
J^a={\cal{T}}_6 ~ j^a$. Since we have chosen massless quarks we use lightcone coordinates for convenience,
\bea\label{lightcone}%
\begin{split}
x^+ = t+z=t-\int d\rho \sqrt{K(r)(R/u)^{3/2}(1+\dot{\sigma}^2)}, \cr
x^- = t-z=t+\int d\rho \sqrt{K(r)(R/u)^{3/2}(1+\dot{\sigma}^2)}.
\end{split}
\eea %
Current conservation condition, $\nabla_a J^a=0$, relates the
currents $j^t$ and $j^{\rho}$ and we demand these currents to be
functions of $x^-(=t-z)$ and define them as derivative of an arbitrary
function $g(x^-)$
\be\label{j-g}%
 j^{\rho}=
-\left[ K(r)(R/u)^{3/2}(1+\dot{\sigma}^2)\right]^{-1/2} j^t=\partial_{-}g(x^-).
\ee %
Making use of \eqref{j-g} in the quark number $\tilde{n}$
relation $\tilde{n}=\int d^6 \xi \sqrt{-g} J^t$, we can determine
the function $g(x^-)$
\be\label{gm} %
g(x^-)\sim\frac{1}{\rm vol(S^2)}
(2\pi)(2\pi\alpha')^3 \lambda n_B(x^-),
\ee %
 where
$n_B(x^-)=\frac{\tilde{n}}{\rm Vol(S^3)}$ is baryon number density.

Planning to have a time dependent chemical potential means that we
should turn on $F_{t\rho}$ component of the time dependent gauge
field in DBI action \eqref{DBI} and add the source term
\eqref{source} to this action, too. The Lagrangian density in the presence of the
magnetic field $B$ is 
\bea\label{B-F-lagrangian} %
\begin{split}
{\cal{L}}_B=\frac{\rho^2 u^{1/4}}{g_s r^2 R^{3/4}} \bigg{\{} \frac{R^{3/2} u^{5/2} [(2\pi\alpha'B)^2 R^3+u^3](1+\dot{\sigma}^2)}{r^2}\cr
-(2\pi\alpha'F_{t\rho})^2(R^{3/2} u^{7/2}+(2\pi\alpha'B)^2 R^{9/2} u)  \bigg{\}}^{1/2}.
\end{split}
\eea %
Now we add the source term \eqref{source} to \eqref{B-F-lagrangian} to find the following DBI action
\bea %
S_{\rm DBI}=-{\cal{T}}_6 \;V_3\; V(\Omega_2) \int dt\; d\rho [{\cal{L}}_B -
(A_t j^t + A_{\rho} j^{\rho})].
\eea %
Equation of motion for the gauge field components 
$A_t$ and $A_{\rho}$ gives following relation for the field strength $F_{t\rho}$ which will be needed in the next subsection,
\bea\label{F-B} %
(2\pi\alpha')F_{t\rho}=\frac{g(x^-) (1+\frac{u_{\rm kk}^3}{4r^3})^{2/3}
\sqrt{1+\dot{\sigma}^2}}{\sqrt{g^2(x^-)+(2\pi\alpha')^2 \rho^4
(1+\frac{u_{\rm kk}^3}{4r^3})^{8/3}[1+((\frac{R}{u})^3(2\pi\alpha'B)^2)]}}.
\eea %

\eqref{F-B} implies that in order to specify $F_{t\rho}$ one needs to determine $g(x^-)$. We pursue the procedure of \cite{Hashimoto:2010wv} and so our baryon
injection model is according to the profile 
\bea\label{g} %
g(x^-)=
\left\{
\begin{split}
	& 0~~~~~~~~~~~~x^-<0\cr
	& g_m \omega x^-~~~~~0<x^-<1/\omega \cr
	& g_m~~~~~~~~~~1/\omega<x^-
\end{split}
\right. .
\eea %
This choice for $g(x^-)$ means that we
start with zero baryon density $n_B = 0$ and then increase it
linearly in time for the interval $0 < t < 1/\omega$. It reaches its maximum value $g_m$ at
$t = 1/\omega$ and then it is kept constant.
Note that $g_m$ in \eqref{g} is given by \eqref{gm} provided that we replace $n_B(x^-)$ with its constant maximum value $n_B$.

\subsection{Fluctuations and Apparent Horizon on the Probe D6-brane}

In the previous subsection we added a time dependent source to DBI action to have a time dependent baryon chemical potential. It finally led to a relation for $F_{t\rho}$, \eqref{F-B}, which depends on time. Presence of a time dependent source on the probe D6-brane modifies
the induced metric. In fact the fluctuations living on the flavor
D6-brane which are gravity dual of mesons feel an effective metric
and appearance of a horizon for this metric corresponds to
thermalization of mesons.

Since we are interested in qualitative behavior of meson thermalization in D4/D6 model and according to
\cite{Hashimoto:2010wv} scalar and vector mesons have similar thermalization time scales, we study scalar fluctuations in this paper.

Suppose $G$ and $g$ as the background and induced metrics
respectively and decompose the induced metric as \bea %
g_{ab}=G_{MN}\partial_{a}X^M \partial_b X^N\equiv
G_{ab}+G_{IJ}\partial_a x^I \partial_b x^J, 
\eea %
where $a,b$ and
$I,J$ indices are respectively used for parallel coordinates with
and transverse coordinates to the worldvolume of D6-brane. Our
goal is finding an effective metric felt by the scalar fluctuations, so
we divide $x^I$ into two parts as
\[
x^I(\xi^a)=\sigma(\rho)+\chi(\xi^a),
\]
where $\sigma(\rho)$ shows the shape of the D6-brane and is a function of
just $\rho$ while $\chi$ exhibits the small fluctuations that can be functions of all worldvolume coordinates $\xi ^a$. To find the
action governing dynamics of mesons in the presence of external fields, we expand the
D6-brane action \eqref{DBI} about the classical solution
$\sigma(\rho)$ for small fluctuations $\chi$, \bea\begin{split} \label{decomposed-action} %
& S_{\rm DBI}=S_0+S_{\rm fluc}=\cr & S_0-\frac{{\cal{T}}_6}{2}\int d^7 \xi
\sqrt{\gamma}\bigg{(}\gamma^{ab}M_{ba}+\gamma^{ab}N_{ba}-\frac{1}{2}\gamma^{ab}M_{bc}
\gamma^{cd}M_{da}+\frac{1}{4}(\gamma^{ab}M_{ba})^2+... \bigg{)},
\end{split}
\eea %
where $\gamma$, $M$ and $N$ are defined as
\bea %
\begin{split}
& \gamma_{ab}=G_{ab}+G_{\sigma\sigma}\partial_a \sigma \partial_b \sigma+(2\pi\alpha')F_{ab},\cr
& M_{ab}=\chi \partial_{\sigma}G_{ab}+G_{\sigma\sigma}\partial_a \sigma \partial_b \chi+G_{\sigma\sigma}\partial_a \chi \partial_b \sigma+\chi \partial_{\sigma} G_{\sigma\sigma} \partial_a \sigma \partial_b \sigma, \cr
& N_{ab}=\frac{1}{2}\chi^2\partial_{\sigma}^2 G_{ab}+G_{\sigma\sigma}\partial_a\chi\partial_b \chi+\chi \partial_{\sigma}G_{\sigma\sigma}\partial_a\sigma\partial_b\chi+\chi\partial_{\sigma}G_{\sigma\sigma}\partial_a\chi\partial_b\sigma+\frac{1}{2}\chi^2\partial_{\sigma}^2 G_{\sigma\sigma}\partial_a\sigma\partial_b\sigma.
\end{split}
\eea %

To calculate the apparent horizon, only the kinetic term of the fluctuations $\chi$ in the action \eqref{decomposed-action} plays role which is given by
\bea\label{kinetic} %
S_{\rm fluc, k}=\frac{-{\cal{T}}_6}{2}\int d^7\xi
\sqrt{\gamma}\;
G_{\sigma\sigma}\left[ S^{ab}(1-G_{\sigma\sigma}\dot{\sigma}^2
S^{\rho\rho})-\dot{\sigma}^2G_{\sigma\sigma}\theta^{a\rho}\theta^{b\rho}\right] \partial_a\chi\partial_b\chi,
\eea %
where $S$ and $\theta$ are the symmetric and antisymmetric
parts of $\gamma$, respectively. Replacing $G_{\sigma \sigma}$ and  
\[
S^{\rho\rho}=\frac{r^2~(u/R)^{3/2}}{u^2(1+\dot{\sigma}^2)-(2\pi\alpha' F_{t\rho})^2 r^2}\;,
\]
in
\eqref{kinetic}, $S_{\rm fluc, k}$ can be written as 
\bea\label{ghat-action} %
S_{\rm fluc,k}=-\frac{{\cal{T}}_6}{2}\int d^7\xi \; A \; G_{\sigma\sigma}
\sqrt{\hat{g}} \hat{g}^{ab} \partial_a \chi \partial_b \chi,
\eea %
where $A=\sqrt{\frac{\gamma}{\hat{g}}}$ and $\hat{g}$ is an
effective metric with components
\bea\label{ghatB} %
{\hat{g}^{ab}}=\frac{(u/r)^2-(2\pi\alpha')^2
F_{t\rho}^2}{(u/r)^2(1+\dot{\sigma}^2)-(2\pi\alpha')^2
F_{t\rho}^2}S^{ab}-\dot{\sigma}^2
G_{\sigma\sigma}\theta^{a\rho}\theta^{b\rho}.\eea

To read apparent horizon we need the action to be written in the form
\bea\label{stilde-action} %
\tilde {S}_{\rm fluc,k}=-\int d^7\xi \;\frac{1}{2}\; \sqrt{\tilde{s}}\; \tilde{s}^{ab}\partial_a\chi\partial_b\chi.
\eea %
Considering $\tilde{s}^{ab}=c\hat{g}^{ab}$ and comparing \eqref{ghat-action} and \eqref{stilde-action} leads to
\[c={\cal{T}}_6^{-2/5}A^{-2/5}G_{\sigma\sigma}^{-2/5}.\]
Therefore the new effective metric $\tilde{s}$ can be obtained from $\hat{g}$. The open string metric components $\tilde{s}_{ab}$ resulting from \eqref{ghatB} in the presence of the magnetic field are
\bea\label{metric-component} %
\begin{split}
\tilde{s}_{tt}=&-{\cal{T}}_6^{2/5}R^{-3/2} u^{7/10} r^{-4/5} (u^3/R^3 +(2\pi\alpha')^2B^2)^{-1/5} (u^2-(2\pi\alpha')^2F_{t\rho}^2 r^2)^{6/5}\\ &(u^2 (1+\dot{\sigma}^2)-(2\pi\alpha')^2 F_{t\rho}^2 r^2)^{-2/5}, \cr
\tilde{s}_{\rho\rho}=&{\cal{T}}_6^{2/5} R^{3/2}u^{-3/10} r^{-14/5}(u^3/R^3 +(2\pi\alpha')^2B^2)^{-1/5} (u^2-(2\pi\alpha')^2F_{t\rho}^2 r^2)^{1/5}\cr &(u^2 (1+\dot{\sigma}^2)-(2\pi\alpha')^2 F_{t\rho}^2 r^2)^{3/5}, \cr
\tilde{s}_{11}=&\tilde{s}_{22}={\cal{T}}_6^{2/5} R^{3/2}u^{-3/10} r^{-4/5} (u^3/R^3 +(2\pi\alpha')^2B^2)^{4/5} (u^2-(2\pi\alpha')^2F_{t\rho}^2 r^2)^{1/5},\cr &(u^2 (1+\dot{\sigma}^2)-(2\pi\alpha')^2 F_{t\rho}^2 r^2)^{-2/5}, \cr
\tilde{s}_{33}=&{\cal{T}}_6^{2/5}R^{-3/2} u^{27/10} r^{-4/5} (u^3/R^3 +(2\pi\alpha')^2B^2)^{-1/5} (u^2-(2\pi\alpha')^2F_{t\rho}^2 r^2)^{1/5}\cr &(u^2 (1+\dot{\sigma}^2)-(2\pi\alpha')^2 F_{t\rho}^2 r^2)^{-2/5}, \cr
\tilde{s}_{\alpha\beta}=&-{\cal{T}}_6^{2/5} R^{3/2} u^{17/10} r^{-14/5}\rho^2 (u^3/R^3 +(2\pi\alpha')^2B^2)^{-1/5} (u^2-(2\pi\alpha')^2F_{t\rho}^2 r^2)^{1/5}\cr &(u^2 (1+\dot{\sigma}^2)-(2\pi\alpha')^2 F_{t\rho}^2 r^2)^{-2/5},
\end{split}
\eea %
where $\alpha$ and $\beta$ stand for angular coordinates of $S^2$ sphere, $\psi_1$ and $\psi_2$, in Table \ref{embedding}.

Given this effective metric we will now determine the apparent horizon.
During blackhole formation one can calculate where, at each instant,
the last outgoing photon will be sent and reach null infinity. This
surface will mark the outermost trapped surface or the apparent
horizon. Finding a condition for apparent horizon formation is
possible by tracking worldline of an outgoing photon. A bunch of
outgoing photons will be trapped if their proper area ($V$) does not
grow in time, $dV/d\tau = 0$ \cite{Luciano}. Therefore we
determine the apparent horizon by demanding that area variation of
the surface 
\bea\label{V} %
V_{\rm surface}=\int d^3x d^2\varphi
\left(\prod_{i=1}^3 \tilde{s}_{ii} \prod_{\alpha=1}^2
\tilde{s}_{\alpha\alpha}\right)^{1/2},
\eea %
vanishes along the null rays normal to the
surface. This is a surface in a given $(t,z)$ at an arbitrary point. $x^i$s ($i=1,2,3$) are
gauge theory directions and $\varphi^\alpha$s ($\alpha=1,2$)
are the angular variables of $S^2$ sphere transverse to D6-brane. The apparent horizon, if
forms, has to satisfy
\bea\label{dv} %
dV_{\rm surface}|_{dt=-dz}=0,
\eea %
that shows a surface whose area is constant along a null ray.
Since the corresponding relation to $V_{\rm surface}$ is too long we do not write it here and deal with \eqref{dv} numerically.

A null vector $(v^a, v^b)$ satisfies the condition $G_{ab} v^a v^b=0$ \cite{Hashimoto:2013mua}. For $(v^t, v^{\rho})$ in the $(t,  \rho)$ slice the
condition of null vector for open string metric
\eqref{metric-component} reduces to
\[
\tilde{s}_{tt} {v^t}^2+\tilde{s}_{\rho \rho} {v^{\rho}}^2=0,
\]
and results in the following ratio
\bea %
\frac{v_t}{v_{\rho}}=R^{3/2} u^{-1/2} r^{-1} \sqrt{\frac{u^2 (1+\dot{\sigma}^2)-(2\pi\alpha')^2 F_{t\rho}^2 r^2}{u^2 -(2\pi\alpha')^2 F_{t\rho}^2 r^2}}.
\eea %
So the apparent horizon formation condition \eqref{dv} becomes
\bea \label{dv-null}%
dV_{\rm surface}=(v^{\rho}\partial_{\rho}+v^t \partial_{t})V_{\rm surface}=0
\eea %

Therefore substituting \eqref{g} in \eqref{F-B} the $\tilde{s}$ components are determined. Then we put them in \eqref{V} to find the surface area $V_{\rm surface}$ and solve \eqref{dv-null} numerically to figure out the apparent horizon.

\section{Numerical Results}

In this section we are going to study meson thermalization by apparent horizon formation on D6-brane, numerically. Supergravity parameters $g_s$, $N_c$, $u_{\rm kk}$ and $R$ are related to $SU(N_c)$
theory parameters $M_{\rm kk}$ and $g_{\rm YM}$ 
according to \eqref{scale}. As \eqref{small-curv} implies
t'Hooft coupling $\lambda$ in this theory is not a fixed parameter
anymore and depends on $u_{\rm kk}$, $\lambda \approx \frac{u_{\rm kk}^{1/2}
R^{3/2}}{l_s^2}$. In addition as was previously
mentioned the curvature of space is proportional to $u_{\rm kk}^{-1/2}$
and so we are not allowed to choose very small values for $u_{\rm kk}$
since it causes very large curvatures. Considering all of these, in this paper we choose $R=1$ and let $u_{\rm kk}$ admit values from $10$ to $100$ that results in
the curvature of order $0.1$ and $\lambda \sim {\cal{O}} (10)$. This
order for $\lambda$ is usually used in gauge/gravity duality
literature. For numerical study of baryon injection we consider the
number of baryons $n_B$ as twice the standard nuclear density $n_N$
times the Lorentz factor $\gamma$, $n_B \sim 2 \gamma n_N$ and for
RHIC data gives $n_B \sim 34$ \cite{Hashimoto:2010wv}. Finally we
set $\omega=1$ in \eqref{g} for numerical computations.

\subsection{Apparent Horizon Formation}
\begin{figure}[ht]
    \begin{center}
        \includegraphics[width=2.6 in]{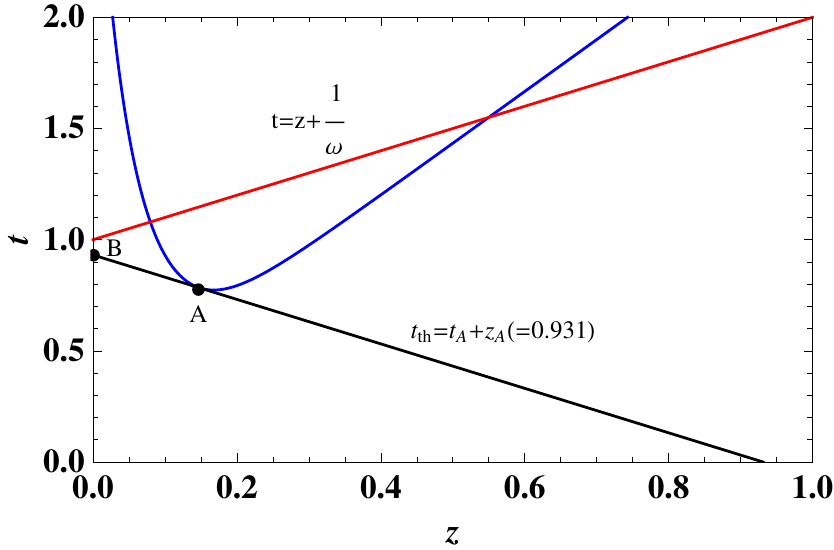}
        \includegraphics[width=2.6 in]{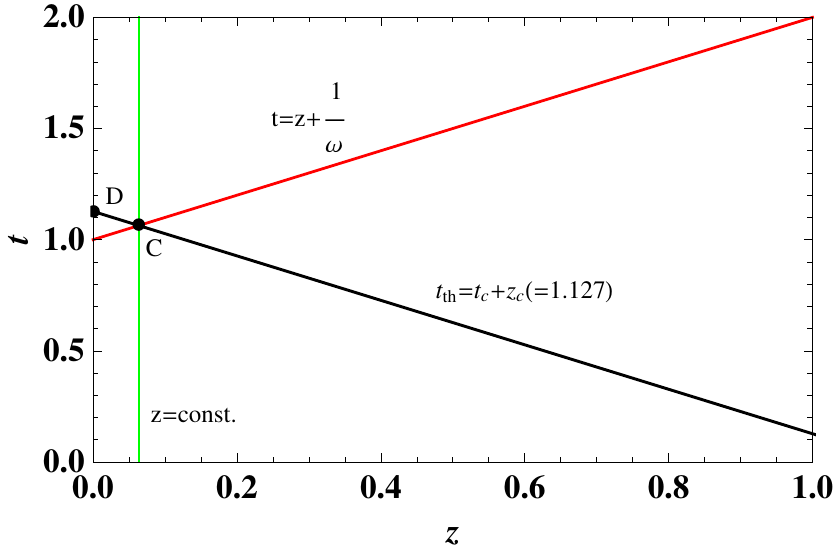}
        \caption{\small Location of the apparent horizon in $z-t$ plane for $g(x^-)=g_m \omega x^-$ (left) and $g=g_m$ (right). Earliest thermalization for the boundary observor occurs at points A (left) and C (right). Black line is the light ray propagating toward boundary.}
        \label{t-z-zero}
    \end{center}
\end{figure}

First we study thermalization for the case $u_{\rm kk}=0$ to have a
qualitative comparison with \cite{Hashimoto:2010wv}, however as we
mentioned earlier it is just a formal discussion. When we choose
$g(x^-)=g_m \omega x^-$ from \eqref{g} and solve \eqref{dv-null},
the blue curve in Fig. \ref{t-z-zero} (left) is obtained. The red
line shows $t=z+1/\omega$ which is plotted to determine the valid
region for blue curve, $t< z+1/\omega$. It means that we are
interested in that part of blue curve which lies under the red line.
The point A at ($t_A , z_A$) in Fig. \ref{t-z-zero} (left)  gives
the earliest delivery of the apparent horizon formation, because the
black line in this figure shows the outgoing null ray toward the
boundary. So the thermalization time seen by the boundary observer
is $t_{\rm th}=t_A + z_A$. This time is equal to $t_B$ since the point
$B$ is located at $z=0$ that determines the boundary.

Now if one chooses $g=g_m$ from \eqref{g} and solve \eqref{dv-null},
the green vertical line in Fig. \ref{t-z-zero} (right) is obtained
and its part above the indicator red line is acceptable according to
its interval in \eqref{g}, $t> z+1/\omega$. In this case the point
$C$ gives the time of the apparent horizon formation that from the
boundary observer point of view occurs at $t_D \equiv t_{th}=t_C
+z_C$. According to the numerical values that we have chosen, it is
seen that $t_B < t_D$ and so the point $A$ gives the earliest
occasion.

\begin{figure}[ht]
    \begin{center}
        \includegraphics[width=2.6 in]{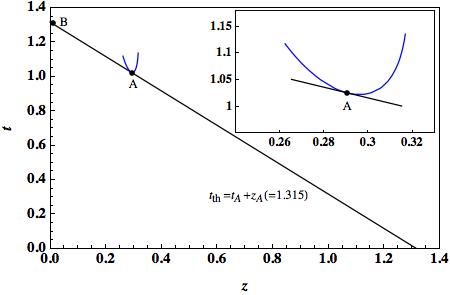}
        \includegraphics[width=2.6 in]{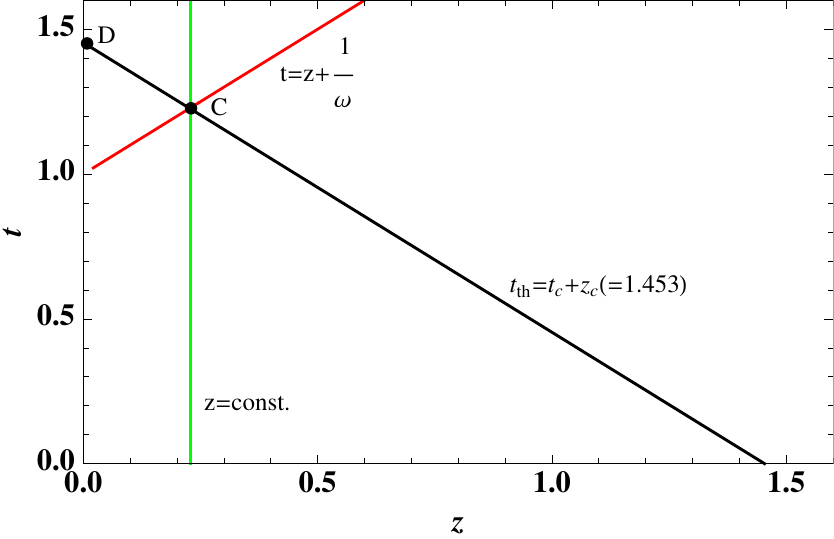}
        \caption{\small Location of the apparent horizon in $z-t$ plane for $g(x^-)=g_m \omega x^-$ (left) and $g=g_m$ (right). $u_{\rm kk}$ is set to $10$. Earliest thermalization for the boundary observor occurs at points A (left) and C (right). Black line is the light ray propagating toward boundary.}
        \label{t-z-uk}
    \end{center}
\end{figure}

Now we study thermalization time in the case of $u_{\rm kk}\neq 0$. For
instance we put $u_{\rm kk}=10$ and by choosing $g(x^-)$ from \eqref{g}
we obtain left and right plots in Fig. \ref{t-z-uk} respectively for
$g(x^-)=g_m \omega x^-$ and $g(x^-)=g_m$. The magnified part in Fig.
\ref{t-z-uk} (left) shows the point of tangency of outgoing null
vector with $t-z$ curve. Again the points B and D show the earliest
times that boundary observer reports horizon formation on D6-brane
for each of the choices of $g(x^-)$. According to selected numerical
values it is seen that the thermalization time measured at B is less
than D and therefore the point A gives the earliest delivery of
information about horizon formation.

All the above discussion and diagrams of apparent horizon formation can be restated in the presence of magnetic field.

\subsection{The Effect of $u_{\rm kk}$ and $B$ on Thermalization Time}

 In this article we defined thermalization as a time-dependent confinment/ deconfinment transition, albeit in meson sector, which occurs due to a sudden change in baryon's number. During this process dynamical degrees of freedom change from mesons to quarks and gluons of thermal plasma. Our goal is to study the effect of scale, $u_{\rm kk}$, and magnetic field, $B$, on transformation of mesons to quarks and gluons in plasma that is addressed in this subsection.

\begin{figure}[ht]
    \begin{center}
        \includegraphics[width=3.6 in]{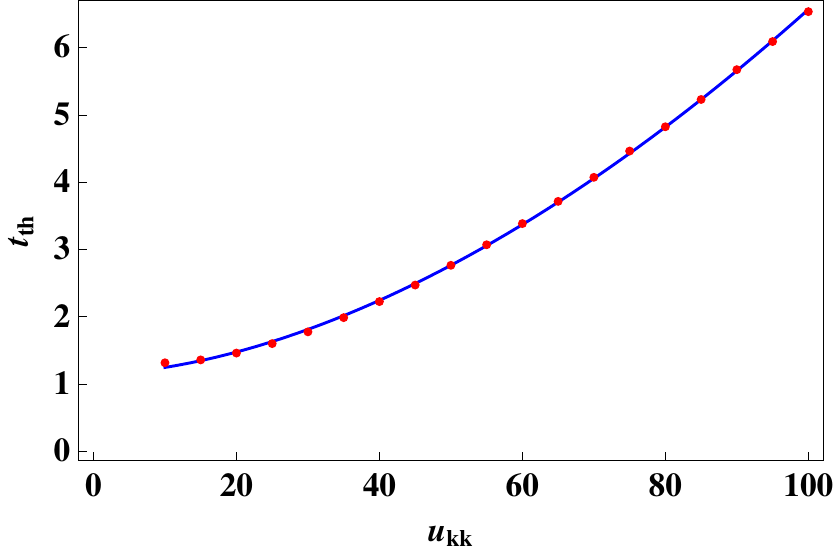}
        \caption{\small Thermalization time for $t<z+1/\omega$ is plotted in terms of $u_{\rm kk}$. The blue curve shows the fitted function, $t_{\rm th}^{u_{\rm kk}}=0.001\;u_{\rm kk}^{1.746}+1.148$. }
        \label{th-uk}
    \end{center}
\end{figure}

In zero magnetic field for different $u_{\rm kk}$'s we obtain
thermalization time according to the procedure of the previous
subsection. Fig. \ref{th-uk} shows the result and it is seen that
thermalization time increases with $u_{\rm kk}$. Recall that we
introduced mesons as fluctuations of the field $\sigma(\rho)$.
According to meson spectroscopy, mass spectrum of these mesons
regardless of the mass of constituent quarks is proportional to the
energy scale of the theory \cite{Kruczenski:2003uq}
\begin{equation}\label{meson-mass}
M_{\rm meson}^2 \sim M_{\rm kk}^2\sim \frac{u_{\rm kk}}{R^3}.
\end{equation}
On the other hand, binding energy of mesons depends on the mass of the meson \cite{CasalderreySolana:2011us}
\begin{equation}\label{binding}
E_B \sim \sqrt{\lambda} M_{\rm meson}.
\end{equation}
Therefore with $u_{\rm kk}$ increment, meson mass and accordingly its
binding energy increase. Since thermalization time is  defined as
the required time for mesons to transform into quark/gluon thermal plasma, so binding energy increase prolongs meson
thermalization (Fig. \ref{th-uk}). We have fitted solid blue curve
\[
t_{\rm th}^{u_{\rm kk}}=0.001\;u_{\rm kk}^{1.746}+1.148,
\]
to data points in Fig. \ref{th-uk}.

\begin{figure}[ht]
    \begin{center}
        \includegraphics[width=3.6 in]{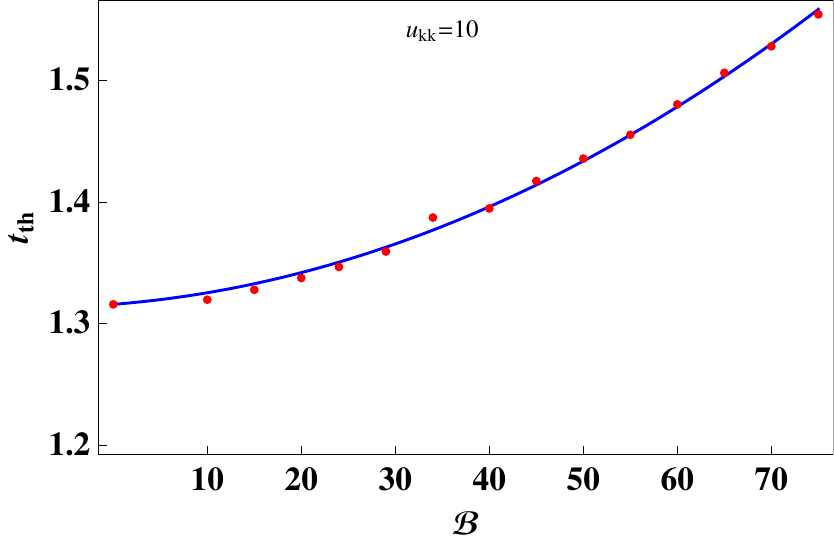}
        \caption{\small Thermalization time for $t<z+1/\omega$ is plotted in terms of  ${\cal B}(=2\pi \alpha' B)$ for a specific value $u_{\rm kk}=10$. The blue curve shows the fitted function, $t_{\rm th}^B=t_{\rm th}^{B=0}(1+4.6 \times 10^{-4} B + 2.6 \times 10^{-6} B^2)$ where $t_{\rm th}^{B=0}=1.3158$.}
        \label{th-b}
    \end{center}
\end{figure}

In the next step for a fixed value of $u_{\rm kk}$ ($u_{\rm kk}=10$) we obtain
thermalization time for different values of the magnetic field. The
result is shown by red points in Fig. \ref{th-b}. As it is seen
thermalization time increases with magnetic field. The reason is the
effect of magnetic field on the binding energy. According
to \cite{Albash:2007bk} magnetic field affects massive mesons by
making them more bounded, and here we are faced with the same
situation because according to \eqref{meson-mass}, $\sigma$ mesons
are massive due to the scale $u_{\rm kk}$. Another study
\cite{Ali-Akbari:2015bha} shows that mass of the meson and therefore
its binding energy increase with magnetic field. So we can conclude that the presence of magnetic field in a system
with energy scale causes meson dissociation to occur later and increases
thermalization time. The blue curve in Fig. \ref{th-b} is a
quadratic function of $B$ which is fitted to data,
\[
t_{\rm th}^B=t_{\rm th}^{B=0}(1+4.6 \times 10^{-4} B + 2.6 \times 10^{-6} B^2).
\]
This qualitative study demonstrates that dependence of $t_{\rm th}$ on magnetic field is insignificant as is pointed in \cite{Fuini:2015hba}.

\section{Summary and Results}

In this article, using AdS/CFT duality, we investigated meson thermalization in QGP in the presence of magnetic field. We supposed that four dimensional large-N$_c$ QCD governs the strongly coupled QGP and we considered a system of N$_c$ D4-branes compactified on a circle as its gravitational dual theory. Inverse of the compactification radius corresponds to energy scale in dual gauge theory and the presence of this scale makes this theory more similar to QCD. To have matter fields we introduced a flavor D6-brane to the system of N$_c$ D4-branes. Then we studied the effect of energy scale and magnetic field on the shape of this flavor brane. As we expected, both quantities prompt a repulsive force between flavor D6-brane and background D4-branes.

Since fluctuations on the flavor brane introduce mesons in the dual gauge theory, to study meson thermalization we first obtained effective metric felt by scalar fluctuations on the brane by expanding DBI action of flavor brane around its classical shape. Then using time dependent components of this metric we wrote the necessary condition for apparent horizon formation on flavor brane which is dual to meson thermalization in dual gauge theory. We considered baryon injection model to mimic the heavy ion collision. It means that we defined baryon-number chemical potential according to a time dependent profile. Using this we introduced the required condition for apparent horizon formation and obtained meson thermalization time in the boundary gauge theory. Finally we studied how the presence of energy scale and magnetic field affect meson thermalization time.

We observed that meson thermalization time or in the other words required time for mesons to transform into quark and gluon degrees of freedom in plasma, increases with energy scale. The reason is that in such a theory, mass of the meson increases when one raises the scale of energy, regardless of the mass of its constituent quarks. This leads to the increase of mesons' binding energy and therefore meson dissociation  becomes harder and thermalization process is prolonged. In addition it was shown that in a fixed energy scale meson thermalization time increases with the magnetic field enhancement. This observation is also justifiable by meson binding energy considerations. In fact it is confirmed in the literature that magnetic field augments the binding energy of mesons and they become more stable. So it is more difficult for them to transform into quark and gluon degrees of freedom in thermal plasma and the thermalization time increases. Note that the effect of magnetic field on thermalization time is not significant except for strong magnetic fields.

\subsection*{Acknowledgments}
The author is thankful to M. Ali-Akbari and M. Rajaee for fruitful comments and discussions.


\begin{thebibliography}{99}

%\cite{Shuryak:2003xe}
\bibitem{Shuryak:2003xe}
E.~Shuryak,
``Why does the quark gluon plasma at RHIC behave as a nearly ideal fluid?'',
Prog.\ Part.\ Nucl.\ Phys.\  {\bf 53}, 273 (2004)
%doi:10.1016/j.ppnp.2004.02.025
[hep-ph/0312227].

%\cite{Shuryak:2004cy}
\bibitem{Shuryak:2004cy}
E.~V.~Shuryak,
``What RHIC experiments and theory tell us about properties of quark-gluon plasma?'',
Nucl.\ Phys.\ A {\bf 750}, 64 (2005)
%doi:10.1016/j.nuclphysa.2004.10.022
[hep-ph/0405066].

%\cite{Heinz:2004pj}
\bibitem{Heinz:2004pj}
U.~W.~Heinz,
``Thermalization at RHIC'',
AIP Conf.\ Proc.\  {\bf 739}, 163 (2005)
%doi:10.1063/1.1843595
[nucl-th/0407067].

%\cite{Luzum:2008cw}
\bibitem{Luzum:2008cw}
M.~Luzum and P.~Romatschke,
``Conformal Relativistic Viscous Hydrodynamics: Applications to RHIC results at s(NN)**(1/2) = 200-GeV'',
Phys.\ Rev.\ C {\bf 78}, 034915 (2008)
%Erratum: [Phys.\ Rev.\ C {\bf 79}, 039903 (2009)]
%doi:10.1103/PhysRevC.78.034915, 10.1103/PhysRevC.79.039903
[arXiv:0804.4015 [nucl-th]].

%\cite{Maldacena:1997re}
\bibitem{Maldacena:1997re}
  J.~M.~Maldacena,
  ``The Large N limit of superconformal field theories and supergravity'',
  Int.\ J.\ Theor.\ Phys.\  {\bf 38}, 1113 (1999)
  %[Adv.\ Theor.\ Math.\ Phys.\  {\bf 2}, 231 (1998)]
  [hep-th/9711200].
  %%CITATION = HEP-TH/9711200;%%
  %10880 citations counted in INSPIRE as of 22 juil. 2015

%\cite{Gubser:1998bc}
\bibitem{Gubser:1998bc}
S.~S.~Gubser, I.~R.~Klebanov and A.~M.~Polyakov,
``Gauge theory correlators from noncritical string theory'',
Phys.\ Lett.\ B {\bf 428}, 105 (1998)
%doi:10.1016/S0370-2693(98)00377-3
[hep-th/9802109].

%\cite{Witten:1998qj}
\bibitem{Witten:1998qj}
E.~Witten,
``Anti-de Sitter space and holography'',
Adv.\ Theor.\ Math.\ Phys.\  {\bf 2}, 253 (1998)
[hep-th/9802150].

\bibitem{CasalderreySolana:2011us}
J. Casalderrey-Solana, H. Liu, D. Mateos, K. Rajagopal and Urs A. Wiedemann,
``Gauge/String Duality, Hot QCD and Heavy Ion Collisions", Cambridge University Press (2014)

%\cite{Witten:1998zw}
\bibitem{Witten:1998zw}
E.~Witten,
``Anti-de Sitter space, thermal phase transition, and confinement in gauge theories'',
Adv.\ Theor.\ Math.\ Phys.\  {\bf 2}, 505 (1998)
[hep-th/9803131].

%\cite{Karch:2002sh}
\bibitem{Karch:2002sh}
A.~Karch and E.~Katz,
``Adding flavor to AdS / CFT'',
JHEP {\bf 0206}, 043 (2002)
%doi:10.1088/1126-6708/2002/06/043
[hep-th/0205236].

\bibitem{Kruczenski:2003uq}
  M.~Kruczenski, D.~Mateos, R.~C.~Myers and D.~J.~Winters,
  ``Towards a holographic dual of large N(c) QCD'',
  JHEP {\bf 0405}, 041 (2004)
  [hep-th/0311270].

%\cite{Ishii:2015gia}
\bibitem{Ishii:2015gia} 
T.~Ishii, E.~Kiritsis and C.~Rosen,
``Thermalization in a Holographic Confining Gauge Theory'',
JHEP {\bf 1508}, 008 (2015)
[arXiv:1503.07766 [hep-th]].

%\cite{Chesler:2008hg}
\bibitem{Chesler:2008hg}
  P.~M.~Chesler and L.~G.~Yaffe,
  ``Horizon formation and far-from-equilibrium isotropization in supersymmetric Yang-Mills plasma'',
  Phys.\ Rev.\ Lett.\  {\bf 102}, 211601 (2009)
 % doi:10.1103/PhysRevLett.102.211601
  [arXiv:0812.2053 [hep-th]].

  %\cite{Chesler:2009cy}
\bibitem{Chesler:2009cy}
  P.~M.~Chesler and L.~G.~Yaffe,
  ``Boost invariant flow, black hole formation, and far-from-equilibrium dynamics in N = 4 supersymmetric Yang-Mills
  theory'',
  Phys.\ Rev.\ D {\bf 82}, 026006 (2010)
%  doi:10.1103/PhysRevD.82.026006
  [arXiv:0906.4426 [hep-th]].

  %\cite{Bhattacharyya:2009uu}
\bibitem{Bhattacharyya:2009uu}
  S.~Bhattacharyya and S.~Minwalla,
  ``Weak Field Black Hole Formation in Asymptotically AdS
  Spacetimes'',
  JHEP {\bf 0909}, 034 (2009)
%  doi:10.1088/1126-6708/2009/09/034
  [arXiv:0904.0464 [hep-th]].

%\cite{Hashimoto:2010wv}
\bibitem{Hashimoto:2010wv}
K.~Hashimoto, N.~Iizuka and T.~Oka,
``Rapid Thermalization by Baryon Injection in Gauge/Gravity Duality'',
Phys.\ Rev.\ D {\bf 84}, 066005 (2011)
[arXiv:1012.4463 [hep-th]].
%%CITATION = ARXIV:1012.4463;%%
%21 citations counted in INSPIRE as of 21 juil. 2015

%\cite{Das:2010yw}
\bibitem{Das:2010yw}
  S.~R.~Das, T.~Nishioka and T.~Takayanagi,
  ``Probe Branes, Time-dependent Couplings and Thermalization in
  AdS/CFT'',
  JHEP {\bf 1007}, 071 (2010)
%  doi:10.1007/JHEP07(2010)071
  [arXiv:1005.3348 [hep-th]].

%\cite{Ali-Akbari:2013tca}
\bibitem{Ali-Akbari:2013tca} 
M.~Ali-Akbari, H.~Ebrahim and Z.~Rezaei,
``Probe Branes Thermalization in External Electric and Magnetic Fields'',
Nucl.\ Phys.\ B {\bf 878}, 150 (2014)
%doi:10.1016/j.nuclphysb.2013.11.012
[arXiv:1307.5629 [hep-th]].

%\cite{AliAkbari:2012hb}
\bibitem{AliAkbari:2012hb}
  M.~Ali-Akbari and H.~Ebrahim,
  ``Meson Thermalization in Various Dimensions'',
  JHEP {\bf 1204}, 145 (2012)
  [arXiv:1203.3425 [hep-th]].
  %%CITATION = ARXIV:1203.3425;%%
  %4 citations counted in INSPIRE as of 25 juil. 2015

%\cite{Kharzeev:2007jp}
\bibitem{Kharzeev:2007jp} 
D.~E.~Kharzeev, L.~D.~McLerran and H.~J.~Warringa,
``The Effects of topological charge change in heavy ion collisions: 'Event by event P and CP violation''',
Nucl.\ Phys.\ A {\bf 803}, 227 (2008)
%doi:10.1016/j.nuclphysa.2008.02.298
[arXiv:0711.0950 [hep-ph]].

%\cite{Bali:2011qj}
\bibitem{Bali:2011qj} 
G.~S.~Bali, F.~Bruckmann, G.~Endrodi, Z.~Fodor, S.~D.~Katz, S.~Krieg, A.~Schafer and K.~K.~Szabo,
``The QCD phase diagram for external magnetic fields'',
JHEP {\bf 1202}, 044 (2012)
%doi:10.1007/JHEP02(2012)044
[arXiv:1111.4956 [hep-lat]].

%\cite{Mamo:2015dea}
\bibitem{Mamo:2015dea} 
K.~A.~Mamo,
``Inverse magnetic catalysis in holographic models of QCD,''
JHEP {\bf 1505}, 121 (2015)
%doi:10.1007/JHEP05(2015)121
[arXiv:1501.03262 [hep-th]].

%\cite{Fukushima:2008xe}
\bibitem{Fukushima:2008xe} 
K.~Fukushima, D.~E.~Kharzeev and H.~J.~Warringa,
``The Chiral Magnetic Effect,''
Phys.\ Rev.\ D {\bf 78}, 074033 (2008)
%doi:10.1103/PhysRevD.78.074033
[arXiv:0808.3382 [hep-ph]].

%\cite{Kharzeev:2010gd}
\bibitem{Kharzeev:2010gd} 
D.~E.~Kharzeev and H.~U.~Yee,
``Chiral Magnetic Wave,''
Phys.\ Rev.\ D {\bf 83}, 085007 (2011)
%doi:10.1103/PhysRevD.83.085007
[arXiv:1012.6026 [hep-th]].

%\cite{AliAkbari:2012vt}
\bibitem{AliAkbari:2012vt} 
M.~Ali-Akbari and H.~Ebrahim,
``Thermalization in External Magnetic Field'',
JHEP {\bf 1303}, 045 (2013)
%doi:10.1007/JHEP03(2013)045
[arXiv:1211.1637 [hep-th]].

%\cite{Mamo:2015aia}
\bibitem{Mamo:2015aia} 
K.~A.~Mamo and H.~U.~Yee,
``Thermalization of Quark-Gluon Plasma in Magnetic Field at Strong Coupling'',
Phys.\ Rev.\ D {\bf 92}, no. 10, 105005 (2015)
%doi:10.1103/PhysRevD.92.105005
[arXiv:1505.01183 [hep-ph]].

%\cite{Fuini:2015hba}
\bibitem{Fuini:2015hba} 
J.~F.~Fuini and L.~G.~Yaffe,
``Far-from-equilibrium dynamics of a strongly coupled non-Abelian plasma with non-zero charge density or external magnetic field'',
JHEP {\bf 1507}, 116 (2015)
%doi:10.1007/JHEP07(2015)116
[arXiv:1503.07148 [hep-th]].

%\cite{Itzhaki:1998dd}
 \bibitem{Itzhaki:1998dd}
 N.~Itzhaki, J.~M.~Maldacena, J.~Sonnenschein and S.~Yankielowicz,
 ``Supergravity and the large N limit of theories with sixteen supercharges'',
  Phys.\ Rev.\ D {\bf 58}, 046004 (1998)
  [hep-th/9802042].

%\cite{Karch:2006bv}
  \bibitem{Karch:2006bv}
  A.~Karch and A.~O'Bannon,
  ``Chiral transition of N=4 super Yang-Mills with flavor on a 3-sphere'',
  Phys.\ Rev.\ D {\bf 74}, 085033 (2006)
 % doi:10.1103/PhysRevD.74.085033
  [hep-th/0605120].

%\cite{Ali-Akbari:2014gia}
\bibitem{Ali-Akbari:2014gia}
M.~Ali-Akbari, Z.~Rezaei and A.~Vahedi,
``Thermal fluctuations and meson melting: a holographic approach'',
J.\ Phys.\ G {\bf 42}, no. 7, 075001 (2015)
%doi:10.1088/0954-3899/42/7/075001
[arXiv:1406.2900 [hep-th]].

\bibitem{Luciano}
 L. Rezzolla, ``An Introduction to Gravitational Collapse to Black Holes'', (2010)

%\cite{Hashimoto:2013mua}
\bibitem{Hashimoto:2013mua}
K.~Hashimoto and T.~Oka,
``Vacuum Instability in Electric Fields via AdS/CFT: Euler-Heisenberg Lagrangian and Planckian Thermalization'',
JHEP {\bf 1310}, 116 (2013)
%doi:10.1007/JHEP10(2013)116
[arXiv:1307.7423].

%\cite{Albash:2007bk}
\bibitem{Albash:2007bk}
T.~Albash, V.~G.~Filev, C.~V.~Johnson and A.~Kundu,
``Finite temperature large N gauge theory with quarks in an external magnetic field'',
JHEP {\bf 0807}, 080 (2008)
%doi:10.1088/1126-6708/2008/07/080
[arXiv:0709.1547 [hep-th]].

%\cite{Ali-Akbari:2015bha}
\bibitem{Ali-Akbari:2015bha}
M.~Ali-Akbari, F.~Charmchi, A.~Davody, H.~Ebrahim and L.~Shahkarami,
``Time-dependent meson melting in an external magnetic field'',
Phys.\ Rev.\ D {\bf 91}, 106008 (2015)
%doi:10.1103/PhysRevD.91.106008
[arXiv:1503.04439 [hep-th]].

\end{thebibliography}
\end{document}